\documentclass[aps,prl,twocolumn,showpacs,superscriptaddress]{revtex4-1}

\usepackage[dvipdfmx]{graphicx}
\usepackage[dvipdfm]{hyperref}
\usepackage{bm}

\begin{document}

\title{Demonstration of displacement sensing of a mg-scale pendulum\\ for mm- and mg- scale gravity measurements}

\author{Nobuyuki Matsumoto}
\email{nobuyuki.matsumoto.e7@tohoku.ac.jp}
\affiliation{Frontier Research Institute for Interdisciplinary Sciences, Tohoku University, Sendai 980-8578, Japan}
\affiliation{Research Institute of Electrical Communication, Tohoku University, Sendai 980-8577, Japan}
\affiliation{JST, PRESTO, Kawaguchi, Saitama 332-0012, Japan}
\author{Masakazu Sugawara}
\affiliation{Research Institute of Electrical Communication, Tohoku University, Sendai 980-8577, Japan}
\author{Seiya Suzuki}
\affiliation{Research Institute of Electrical Communication, Tohoku University, Sendai 980-8577, Japan}
\author{Naofumi Abe}
\affiliation{Research Institute of Electrical Communication, Tohoku University, Sendai 980-8577, Japan}
\author{Kentaro Komori}
\affiliation{Department of Physics, University of Tokyo, Bunkyo, Tokyo 113-0033, Japan}
\author{Yuta Michimura}
\affiliation{Department of Physics, University of Tokyo, Bunkyo, Tokyo 113-0033, Japan}
\author{Yoichi Aso}
 \affiliation{National Astronomical Observatory of Japan, Mitaka, Tokyo 181-8588, Japan}
 \affiliation{Department of Astronomical Science,  SOKENDAI (The Graduate University
for Advanced Studies), Mitaka, Tokyo 181-8588, Japan}
\author{Seth B. Cata\~{n}o-Lopez}
\affiliation{Research Institute of Electrical Communication, Tohoku University, Sendai 980-8577, Japan}
\author{Keiichi Edamatsu}
\affiliation{Research Institute of Electrical Communication, Tohoku University, Sendai 980-8577, Japan}
\date{\today}

\begin{abstract}
Gravity generated by large masses has been observed using a variety of probes  from atomic interferometers to torsional balances. 
However, gravitational coupling between small masses has never been observed so far. 
Here, we demonstrate sensitive displacement sensing of the Brownian motion of an optically trapped 7-mg pendulum motion whose natural quality factor is increased to $10^8$ 
through dissipation dilution. 
The sensitivity for an integration time of one second corresponds to the displacement generated by the gravitational coupling between the probe and a mm separated 100 mg mass, whose position is modulated at the pendulum mechanical resonant frequency. 
Development of such a sensitive displacement sensor using a mg-scale device will pave the way for a new class of experiments where gravitational coupling between small masses in quantum regimes can be achieved.
\end{abstract}


\maketitle

{\it Introduction.}---
Ever since the pioneering works of Maskelyne \cite{maskelyne} and Cavendish \cite{cavendish}, gravitational coupling has been explored in connection with, e.g. the gravitational constant \cite{Grev}, gravitational waves \cite{ligo1,ligo2}, gravity-induced decoherence \cite{bassi}, and extra dimensions \cite{murata}. 
In addition to these interests, gravitational coupling between masses in quantum regimes has been attracting attention as a possible key to the experimental exploration of quantum gravity \cite{dewitt,marletto,bose,belenchia,dan}. 
However, in general, while quantum control has been realized for small and hence low-mass oscillators, gravity has only been observed using relatively large masses.

On one hand, the recently developing field of cavity-optomechanics \cite{aspelmeyer} is gradually changing the situation using a bottom-up approach, i.e. increasing the mass scale for which the quantum ground state \cite{chan,teufel,peterson} and quantum control can be realized \cite{riedinger,korppi}. 
Since quantum states easily decohere due to coupling to the environment, e.g. due to black-body radiation \cite{joos,schlosshauer}, this bottom-up approach is limited by the scale of  the de Broglie wavelength of thermal photons of mm-scale, giving an associated mass scale for quantum oscillators of $\sim$ mg. 
On the other hand, the recently proposed protocol given in Ref.\ \cite{schmole} changes the situation to one of a top-down approach, i.e.,\ observation of gravity between small masses.  
Contrary to the conventional approach to such gravity measurements, the proposed protocol is based on the sensitive displacement measurement of a probe oscillator, which is driven by modulated gravity. 
Because the modulation improves the signal-to-noise ratio, sensitivity can be improved by up to at least three orders of magnitude relative to current state-of-the-art experiments with source masses of 100 g \cite{gillies,rogers}.   
Thus, it is envisioned that the progress of both approaches will lead to quantum experiments on mm- and mg- scale oscillators for which sensitive gravity measurements may also be made. 

A few experiments involving displacement sensing of a mg-scale oscillator have been carried out so far in the field of the gravitational-wave detectors \cite{matsumoto2014,matsumoto2015,matsumoto2016}. 
However, to the best of our knowledge, no experiment exists which has sufficient sensitivity to resolve the effect of gravity of the mg- and mm- scale. 
In order to evaluate the sensitivity necessary to resolve the effect of gravity in this regime, we calculate the root mean squared value of the displacement due to gravitational interaction which is given by \cite{schmole}
\begin{eqnarray}
\sqrt{P_{xxG}}&=&\sqrt{2\pi}\frac{Q_{\rm m}}{\omega_{\rm m}^2}GM\frac{d_s}{d_0^3}\nonumber\\
&\simeq&3\times10^{-14}\ [{\rm{m}}] \frac{Q_{\rm m}}{250}\frac{M}{100\ [{\rm{mg}}]}\left(\frac{280\ [{\rm{Hz}}]}{\omega_{\rm m}/2\pi}\right)^2, 
\label{eq}
\end{eqnarray}
where $Q_{\rm m}$ is the quality factor of the mechanical oscillator, $\omega_{\rm m}/2\pi$ is the mechanical resonant frequency, $G$ is the gravitational constant, $M$ is the mass of the source mass, $d_0$ is the separation between the source mass and the probe mass, and $d_s$ is the driving amplitude of the source mass. 
Here, we suppose a  gravity source of 100 mg, separated by 3.75 mm and having a 1.25 mm driving amplitude as in Ref.\ \cite{schmole}. 
\begin{figure*}
\centering
\includegraphics[width=170mm]{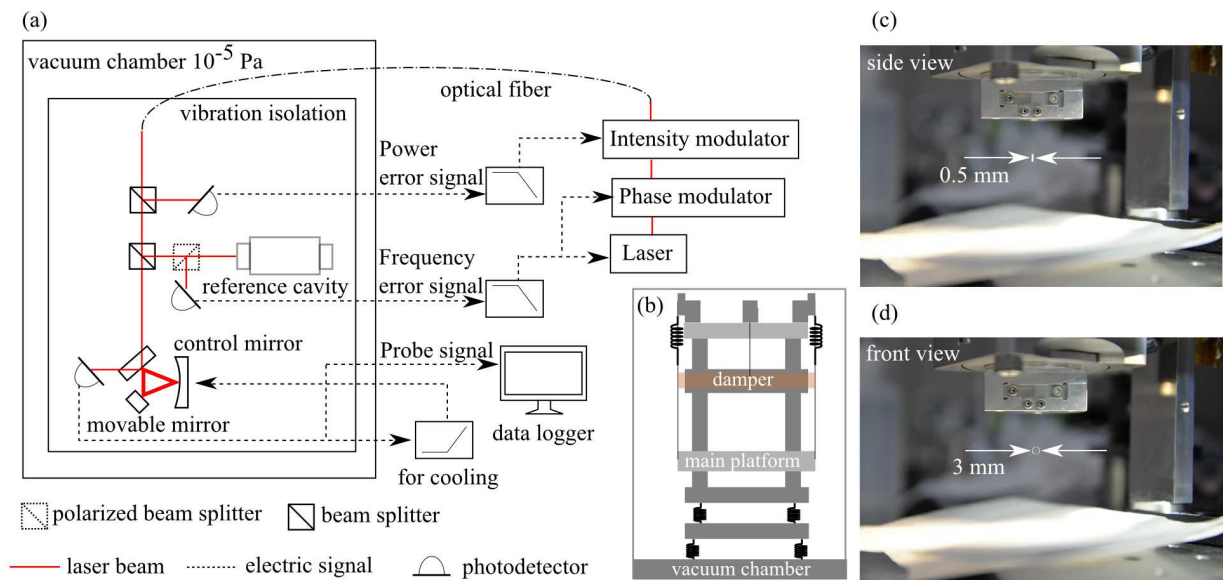}
\caption{(Color online) (a) Experimental setup. 
Outside the vacuum system, a phase modulator and an intensity modulator are inserted between the optical fiber and the laser source for laser stabilization. 
Inside the vacuum chamber, a five stage vibration isolation setup is installed with the monitor for laser stabilization and the ring cavity being fixed to the maximally isolated stage. 
(b) Vibration isolation system. 
Inside the vacuum chamber, we install the five stage vibration isolation stage. 
On the main platform, the laser intensity monitor, the reference cavity, and the ring cavity are installed. 
(c), (d) Mechanical oscillator. 
The pictures show the 7 mg movable mirror of 3 mm diameter and 0.5 mm thickness, which is photographed before setting it on the main plate as shown in Fig. 1 (b). 
}
\label{fig1}
\end{figure*}
Thus, fixing the mechanical resonance frequency at 280 Hz (which is the frequency applicable in the experiment we will consider below), any experiment to measure gravity in the crucial mg- and mm- regime must have a sensitivity of at least $3\times 10^{-14} {\rm m/\sqrt{Hz}}$. 

In this Letter, we report on the first realization of displacement sensing of a mg-scale probe pendulum which meets the requirement of Eq.\ (\ref{eq}) for an integration time of one second. 
Our displacement sensor is an optical ring cavity with a movable planar end mirror, which was initially constructed for the purpose of realizing active feedback cooling of a movable mirror \cite{matsumoto2016}, based on the technology of gravitational-wave detectors \cite{matsumoto2014}. 
All of the applications for active feedback cooling, measuring gravity, and measuring gravitational-waves rely on sensitive displacement measurements, and hence our approach is a top-down approach for both gravity measurements and quantum experiments.

{\it Experiment.}---
The experimental setup shown in Figs.\ \ref{fig1} (a) and (b) was used to perform sensitive displacement measurements of the 7 mg movable mirror shown in Figs. 1(c) and (d). 
For the displacement sensor, we used a triangular optical cavity (to stabilize the yaw mode \cite{matsumoto2014}) of finesse 1800 and 10 cm length, which consists of the movable 7 mg mirror, a fixed mirror, and a 300 g suspended mirror attached to a coil-magnet actuator for cavity length control. 
The position of the 300 g mirror is locally controlled using a Michelson interferometer (omitted in Fig.\ \ref{fig1} (a) for simplicity) in order to reduce the fluctuation of the displacement. 
The mg-mirror is suspended as a pendulum by a 1 cm long silica fiber of 1 $\mu$m diameter, whose material quality factor is 1000 as measured by a ringdown measurement of the pendulum yaw mode. 
The fiber is manufactured using the heat-and-pull method \cite{birks} using a tapering rig developed in \cite{mark}.
The natural oscillation frequency of the pendulum (i.e. the center of mass (COM) mode) $\omega_0/2\pi$ is 4.4 Hz and the COM's natural quality factor $Q_0$ is $10^5$ as measured by a ringdown measurement of the pendulum oscillatory mode. 
The optical cavity is driven by a stabilized continuous-wave laser of 1064 nm wavelength and 30 mW input power. 
The cavity length is controlled to be slightly larger than the optical resonant condition so that a positive restoring force is imposed on the movable mirror due to radiation pressure. 
This optical spring effect increases the movable mirror's COM mode to 280 Hz and also increases the natural quality factor of the COM mode to $10^8$. 
Based on the analysis of transfer functions of the optomechanical system as in Ref.\ \cite{matsumoto2016}, we obtained a voltage-to-displacement conversion factor of $2\times10^{-10}$\ m/V, allowing us to analyze the displacement of the movable mirror by monitoring the intensity of the cavity output.  
In addition, we apply active feedback cooling to the system in order to prevent instability due to the negative value of the optically trapped COM's dissipation rate. 
The quality factor is controlled to be 250. 
All of the ring cavity and the laser stabilization system except for the intensity and phase modulators are set inside a vacuum chamber of $10^{-5}$ Pa. 
The magnitude of this pressure corresponds to the COM's dissipation rate of a few nano Hz so that the quality factor would be limited to around $10^9\times\omega_{\rm m}$ due to residual gas damping. 

\begin{figure*}
\includegraphics[width=181mm]{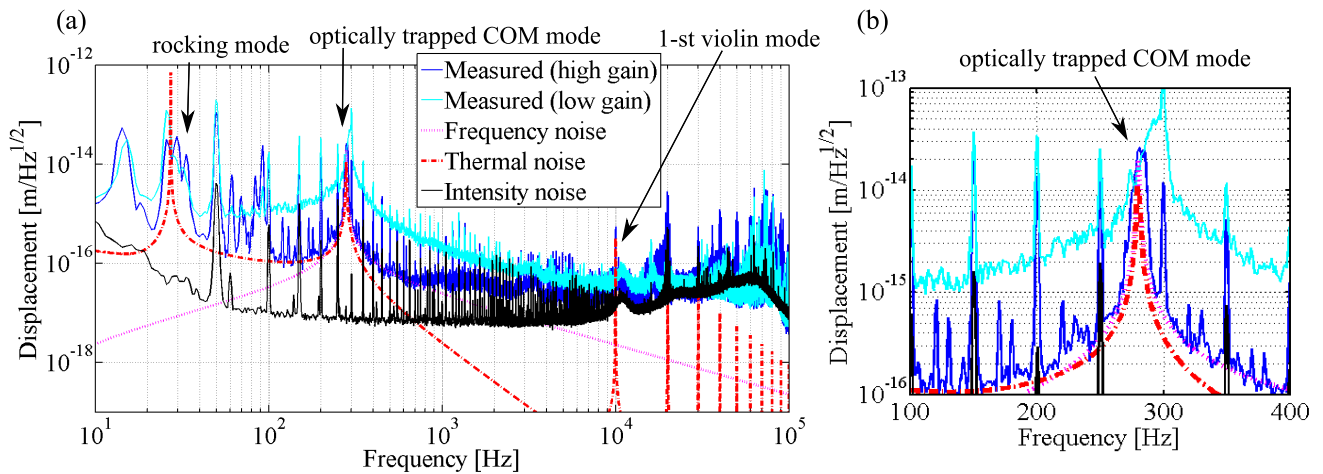}
\caption{(Color online) (a) The measured spectra with the bandwidth of 1 Hz and the noise budget from 10 Hz to 100 kHz. 
The suspension thermal noise is calculated using the result in \cite{gonzalez} and the frequency noise modified by the optical spring effect is calculated using the result in \cite{matsumoto2016}. 
(b) The spectra from 100 to 400 Hz. 
By combining the thermal noise and the frequency noise due to the optical spring, the width of the spectral peak at the resonant frequency of 280 Hz is broadened. 
}
\label{fig2}
\end{figure*}

We use a laser-diode-pumped single-mode Nd:YAG laser (Coherent, Mephisto 500). 
A fraction of the beam is introduced into the vacuum chamber through a polarization maintaining fiber. 
Inside the vacuum chamber, the light is divided and sent to a photodetector (HAMAMATSU, G10899-03K, InGaAs photodiode) to stabilize the laser power, a reference cavity to stabilize the laser frequency, and to the ring cavity itself. 
Since the beam hits each photodetector at an incidence angle of 45, it is possible to absorb residual reflections at each photodiode surface with absorption filters (Thorlabs, NENIR260B). 
The wavelength of the laser is actively stabilized against a reference cavity (Crystalline Mirror Solutions, 300 mm round-trip length) of finesse 150000 by servo loops with a unity gain frequency of 100 kHz. 

In order to reject seismic noise, all of the sensors for the laser stabilization and the displacement measurement are installed on a five-stage vibration isolated aluminum plate inside a vacuum chamber of 1 m diameter and 1 m height  as shown in Fig.\ 1(b). 
The vibration isolation system consists of double pendulums, a vertical spring and double isolation stacks. 
The double pendulum configuration is as follows: the experiment base plate of 750 mm diameter was suspended from an intermediate damping mass made of a copper ring of diameter 750 mm and radial thickness 30 mm using three tungsten wires of diameter 0.45 mm. Strong eddy-current damping of the intermediate mass' motion is realized by surrounding the inside by another suspended ring mass made of magnetic stainless steel and installing 60 neodynium magnets on the intermediate mass' rim.

Fine alignment of the optical systems installed within the vacuum chamber is achieved by picomotor actuators (Newport 8353V, 8301V) which are controlled remotely by driving signals transferred through thin copper wires of diameter 50 or 100 $\mu$m.
We note that we originally used wires of 250 $\mu$m diameter for this purpose, but this was found to lead to contamination of the noise spectral density of the position measurement.

By comparison with our previous report \cite{matsumoto2016}, in the present experiment: (i) the suspension wire for the mg-scale mirror was changed from a tungsten wire of 5 cm length and 3 $\mu$m diameter to a silica wire of 1 cm length and 1 $\mu$m diameter in order to increase the resonant frequency of the violin modes from 190 Hz to 10 kHz; (ii) the pressure inside the vacuum chamber was reduced to $10^{-5}$ Pa by changing the turbo-molecular pump to one with a high pumping speed of 2200 L/s (Edwards, STP-iXA2205CP) and changing almost all vacuum incompatible devices to vacuum compatible ones, in order to reduce the residual gas damping rate; (iii) the laser frequency is stabilized using a reference cavity installed on the vibration isolated stage; (iv) the local control of the position of the mirror attached to the actuator is installed on the same vibration isolated plate; and (v) the vacuum chamber is changed from one of 50 cm diameter to one of 100 cm diameter in order to make the all-in-vacuum and all-on-vibration-isolated system in order to achieve a small susceptibility to environmental noise.  
After these improvements, the ring cavity had sufficient displacement sensitivity to resolve gravity of mg- and mm- scale. 

{\it Data analysis and discussions.}---
The measured power spectrum of the mirror displacement is shown in Figs.\ \ref{fig2} (a) and (b) for two different laser frequency noise levels as achieved by changing the feedback gain. 
The data for the two displacement spectra and the intensity noise spectrum were not measured at the same time.  
As shown in Fig.\ 2\ (b), the data shows that the noise level around the measured COM resonance at low feedback gain originates in the fluctuations of the laser frequency. 
In the data for the high feedback gain, the suspension thermal noise with the structural damping model \cite{gonzalez} (i.e. the dissipation rate being inversely proportional to the frequency) agrees well with the result below the resonance. 
Because the residual gas damping can be ignored, and the pendulum's potential consists of three contributions from the {\it lossless} gravitational potential, the {\it lossless} optical potential, and the dissipative {\it weak} elastic potential, the structural damping model can be applied.  
Due to the frequency dependence of the structural damping, the natural quality factor of the COM at 280 Hz reaches $1\times 10^8$, which is comparable to the reported highest quality factor in Ref.\ \cite{niles}. 
Note that this value of the quality factor is smaller than the ideal one given by $Q_0(\omega_{\rm m}/\omega_0)^2$ by a factor of 4 because a pendulum with a single wire always induces coupling between the COM and the slightly dissipative pitching mode which results in mode mixing between them and the reduction of the quality factor. 
In addition, the predictions of the structural damping model provide a check of the validity of the displacement calibration, because the measured data shows good agreement with the calculated thermal noised spectrum, which is obtained by using independently measured values of the mass of the mechanical oscillator and the natural quality factor $Q_0$. 
Note that harmonics of the power supply frequency of 50 Hz are visible in the spectrum. These harmonics persisted even when we transferred our measurement system electronics to battery power. This suggests that the use of coaxial cables to transfer the signals is the cause of this contamination.

The peak sensitivity is $3\times10^{-14}$ m/$\sqrt{\rm{Hz}}$ at 280 Hz with a quality factor of 250. To the best of our knowledge, this is the world-record sensitivity level achieved for a mg-scale probe. Furthermore, this sensitivity also meets the requirements of Eq.\ ({\ref{eq}}) for an integration time of only one second, meaning that our pendulum displacement measurements are sufficiently sensitive to probe gravity in the crucial mm- mg- regime. 
Note that although we apply active feedback cooling to the system, this does not change the signal to noise ratio between  gravity and either the thermal noise, or the optical spring modified frequency noise, while it greatly eases the requirements of the measurement time for resolving the mechanical motion (i.e., measurement time should be much larger than the mechanical decay rate).

Additionally, we expect that our system can potentially realize cooling of a mg-scale oscillator to an effective temperature of a few $\mu$K, where the effective temperature is given by $4T(1/Q_0)(\omega_0/\omega_{\rm m})^2$. 
Here $T$ is the room temperature and the factor of 4 is due to the mode mixing. 
We note that the effective temperature is proportional to the square of $(\omega_0/\omega_{\rm m})$ while it is linearly proportional to the factor when the viscous damping model is considered. 

The ground state of the mg-scale oscillator can be realized by (i) increasing the COM's resonance to 2 kHz; (ii)  reducing the laser noise; and (iii) reducing the coating thermal noise. 
The increase of the resonant frequency improves the $Qf$ product from $3\times10^{10}$ to $10^{13}$, which means the number of coherent oscillation can exceed one. 
The frequency noise can be reduced by reducing the cavity length by changing the ring cavity with the planar 7 mg mirror to a linear cavity with a concave 7 mg mirror, as has recently been developed by the OptoSigma company. 
Since a linear cavity within a negative-g parameter regime is stable for the mirror's yaw mode, as it is in the case of the present triangular cavity \cite{matsumoto2014}, it is possible to reduce the cavity length while maintaining the stability.  
The intensity noise can be reduced by increasing the finesse. 
The coating thermal noise should be reduced using a crystalline coating \cite{cole} with the coherent cancellation technique \cite{rana}. 
With these improvements, it is possible to meet the requirement of ground state cooling, that is, to achieve a time scale for resolving the zero point motion of the movable mirror faster than the rethermalization rate, which was recently realized in Ref.\ \cite{kippen}. 


Throughout this paper, we assume and show data for a measurement bandwidth of 1 Hz. 
Although our optical cavity can maintain the resonant condition for up to one hour under the high vacuum condition, the data were recorded with a sampling frequency of 1 MHz for ten seconds due to the memory limitation of our data logger (YOKOGAWA, DL850). 
In principle, the measurement bandwidth for our system can be decreased to at least 0.01 Hz, and hence the stochastic noise power spectral density could be reduced by a factor of 100. 
Thus, it is possible to exploit the sensitivity of a few mg scale devices to measure the gravitational force exerted by a mass of the same size separated by 4 mm in distance. 

Lastly, although the proposed probe for measuring gravity was a cantilever in Ref.\ \cite{schmole}, in the present experiment, we use a pendulum as a probe.  
Because the pendulum's oscillation frequency can be controlled by the optical spring effect, the use of a pendulum can allow a new approach to discriminate between the gravitational signal and noise synchronized with the driving of a source mass, in particular, by measuring the dependence of the spectrum on the optical rigidity. 
The ability to tune the resonant frequency will be useful for future gravity measurements based on the modulation technique of Ref.\ \cite{schmole}. 

{\it Conclusion.}---
We realized sensitive displacement sensing using a mg-scale probe mass based on the dissipation dilution technique. Our probe could be used to measure the gravitational coupling between mg-scale masses. 
The sensitivity presented here corresponds to the displacement generated by a driven 100 mg source mass, which also shows that the natural quality factor of the optically trapped pendulum of the probe mass reaches $10^8$. 
We believe our results not only represent an important step towards the measurement of gravity between small masses, but also an important step towards the ground state cooling of a mg-scale oscillator.  

{\it Acknowledgment.}--- 
We thank Mark Sadgrove for use of the fiber tapering rig, help with the manuscript and stimulating discussions. 
We thank Markus Aspelmeyer, Tobias Westphal, David Moore, Jason Twamley, Matthew Obrien, Sougato Bose, Matthew Flinders, and Daniel Carney for fruitful discussions. 
N. M. is supported by PRESTO, JST, JSPS KAKENHI Grant No. 15617498 and Matsuo Academic Foundation.
Y. A. and N. M. were supported by NINS Program for Cross-Disciplinary Study.   

{\it Contribution.}--- 
N. M. designed and planed this research. 
N. M. carried out the measurements, analysis and wrote the paper. 
M. S. and N. M. made the silica fiber. 
N. M., S. S., N. A., and S. B. C. L. assembled the vibration isolation system. 
N. M., N. A., S. S., and K. K. assembled the optical system. 
All authors discussed the results and commented on the manuscript. 

\end{document}